\documentclass[a4paper,11pt]{article}
\pdfoutput=1  
\usepackage{jcappub}
\usepackage{graphicx}
\usepackage{subfig}
\graphicspath{{./figures/}}
\definecolor{MyBlue}{rgb}{0.15,0.15,0.70}
\definecolor{MyGreen}{rgb}{0.15,0.70,0.15}
\newcommand{\beqa}{\begin{eqnarray}}
\newcommand{\eeqa}{\end{eqnarray}}
\hypersetup{
colorlinks=true,
citecolor=MyBlue,
linkcolor=MyBlue,
urlcolor=MyBlue
}

\newcommand{\troisj}[6]{\left(\begin{array}{ccc}
      #1 & #2 & #3 \\
      #4 & #5 & #6\end{array}\right)}

\usepackage{amssymb}
\usepackage{amsmath}
\usepackage{amsfonts}
\usepackage{upgreek}
\usepackage{latexsym}
\usepackage{stfloats}
\usepackage{afterpage}

\newcommand{\be}{\begin{equation}}
\newcommand{\ee}{\end{equation}}
\newcommand{\beq}{\begin{equation}}
\newcommand{\eeq}{\end{equation}}
\newcommand{\bea}{\begin{eqnarray}}
\newcommand{\eea}{\end{eqnarray}}

\newcommand{\bv}{{\bf{v}}}

\newcommand{\bm}{{\bf{}}}
\newcommand{\dd}{\text{d}}
\newcommand{\bx}{{\bf{x}}}
\newcommand{\bn}{{\bf{n}}}

\newcommand{\nn}{{\nonumber}}

\usepackage[stable]{footmisc}

\newcommand\ees{\end{eqnarray}}
\newcommand\bees{\begin{eqnarray}}

\title{CMB  sky  for an off-center observer in a local void I: framework for forecasts}

\author[a,b]{Viraj Nistane}
\author[c]{Giulia Cusin}
\author[a]{Martin Kunz}

\affiliation[a]{
Universit\'e de Gen\`eve, D\'epartement de Physique Th\'eorique and Centre for Astroparticle Physics,
24 quai Ernest-Ansermet, CH-1211 Gen\`eve 4, Switzerland
}
\affiliation[b]{Universit\"ats-Sternwarte M\"unchen (Ludwig-Maximilian-Universit\"at), Scheinerstra\ss e 1, D-81679 M\"unchen, Germany}
\affiliation[c]{Astrophysics  Department,  University  of  Oxford,  DWB,  Keble  Road,  Oxford  OX1  3RH,  UK}

\emailAdd{viraj.nistane@unige.ch}
\emailAdd{giulia.cusin@physics.ox.ac.uk}
\emailAdd{martin.kunz@unige.ch}

\abstract{
The Universe is not perfectly homogeneous, the large scale structure forms overdense regions and voids. In this paper, we consider the possibility that we occupy a special position in our Universe, close to the center of a local underdense region that we model as an LTB void embedded in a homogeneous and isotropic Universe. The CMB sky measured by an off-center observer in this void is not statistically isotropic. In addition to the non-stochastic CMB anisotropies due to the geometry of the model, we also observe a lensing-like distortion of the CMB anisotropies. In this article, we propose a framework to forecast the precision with which we can measure the amplitude of the lensing-like deformation of the CMB temperature anisotropies. For illustrative purposes, we apply this method to a couple of large-scale void models differing over the matter density profile and we show that the CMB temperature data from the Planck satellite is potentially capable to detect the effect for the large voids chosen here.  A companion paper will be dedicated to a systematic exploration of different realistic void models.
}

\begin{document}

\maketitle




\section{Introduction}

In the standard lore of the construction of a cosmological model  the universe on large scale is assumed to be spatially homogeneous and isotropic. In such a universe, an observer in any position would observe exactly the same Cosmic Microwave Background (CMB) sky.  In this work we explore the possibility that we occupy a privileged position, inside a local underdense region. Inside such a local void, unless the observer is positioned exactly at the center of the bubble, the distribution of matter, as seen by the observer, will be anisotropic. This will affect the observed microwave background (and other cosmological observables).

Specific models that give rise to such underdensities have been studied  in the form of a local homogeneous void \cite{Tomita:1999qn, Tomita:2001gh, Tomita:2000jj}. In these works both the underdensity and the region outside it are assumed to be perfectly homogeneous Friedmann-Lema\^\i tre-Robertson-Walker (FLRW) models with a singular mass shell separating the two regions. The inhomogeneity manifests itself as a discontinuity at the location of the mass shell.
In this article, we investigate a more realistic model where the transition between the inner underdensity and the outer region is continuous. Specifically,  we consider an isotropic but inhomogeneous matter-dominated universe model, where the inhomogeneity is spherically symmetric. The model can then be described within the Lema\^\i tre-Tolman-Bondi (LTB) class of spherically symmetric universe models \cite{1933ASSB...53...51L, 1934PNAS...20..169T, Bondi:1947fta}. To make contact with the ordinary FLRW models, we assume that the universe is homogeneous except for an isotropic inhomogeneity of limited spatial extension, with the transition between these two regions being continuous. We consider the observer displaced with respect to the center of this local void. A similar setting has been recently considered in \cite{Alnes:2006pf, Alnes:2005rw, Fanizza:2014baa}. However, our goal is not to propose such a void as a way to circumvent the need for dark energy. Instead, we are interested in the impact of voids, or more generally local inhomogeneities, on the CMB. 

We compute the temperature anisotropies observed by the observer, distinguishing different contributions: primary, geometry-induced non-stochastic and lensing-induced anisotropies. As recently studied in \cite{Cusin:2016kqx} at first order in the lensing potential, all multipole coefficients of the temperature anisotropies are then correlated. In contrast, the correlation matrix measured by a boosted observer in a homogeneous and isotropic universe at linear order in the boost velocity has a simple structure, with non-vanishing correlation only among multipoles separated by $\pm 1$. This is explained in some details later in Sect.\ \ref{correlation}. As observed in \cite{Cusin:2016kqx} this an interesting signature of the void model, which in principle would allow one to distinguish geometrical effects from kinematical effects in CMB observations.

In this paper we consider for the first time the structure of these off-diagonal terms for a general void given by an LTB metric.
We split the theoretical correlation function into a contribution from primary anisotropies which contains only diagonal elements and one from lensing which contains also the off-diagonal terms. We forecast the precision with which we can measure the amplitude of the lensing contribution. To this goal, we keep all cosmological and void/observer parameters fixed (parameter degeneracies can increase the error bars on  the amplitude).
To illustrate the method, we apply it to two large void models with different matter density profiles and we compute the Signal-to-Noise ratio for both cases. 
We choose the void models proposed in \cite{Alnes:2005rw,Alnes:2006pf} for historical reasons and because these models, being relatively simple, allow us to have a pedagogical presentation of our method. \footnote{We stress that the models in \cite{Alnes:2005rw,Alnes:2006pf} have been shown to be ruled out for the purpose of serving as an alternative to dark energy \cite{Moss:2011ze,Zhang:2010fa,Bull:2011wi,Chluba:2014wda}. Some previous works describing more realistic void models (LTB with cosmological constant) are \cite{Valkenburg:2012td,Valkenburg:2011tm,Valkenburg:2011ty,Romano:2010nc}.}
We find that these specific examples would leave a detectable signal in the Planck\footnote{\url{http://sci.esa.int/planck}} data, but a more extensive investigation is deferred to a companion paper \cite{viraj2} dedicated to applying this framework systematically to a large class of void models with different void sizes, observer positions and matter profiles.

We note that the approach that we use to characterize the CMB temperature correlation function in the void model captures an arbitrary angular dependence of this observable. Similar approaches has been used in the past in other contexts, to analyse CMB~\cite{Cusin:2016kqx, Hajian:2003qq,Hajian:2005jh,Basak:2006ew,Pullen:2007tu,  2012PhRvD..85b3010B} and LSS~\cite{Heavens:1994iq,Hamilton:1995px,Szalay:1997cc,Szapudi:2004gh,Papai:2008bd,Bertacca:2012tp,Shiraishi:2016wec,Bartolo:2017sbu,Sugiyama:2017ggb,Tansella:2018hdm} data.

This article is organized as follows. The description of the void model we use is presented in Section \ref{model}. In Section \ref{CMB} we derive the expression for temperature anisotropies distinguishing different contributions and in Section \ref{correlation} we derive the corresponding correlation function for the lensing induced component. We consider additionally the possibility that the observer has a peculiar motion with respect to the CMB rest frame and we review the implication of the at the level of the temperature correlation function. In Section \ref{results} we present a framework to study how well the amplitude of the statistically anisotropic part of the correlation function can be constrained and we apply it to study the Signal-to-Noise for a couple of realistic void models.


\section{General framework}\label{model}

We want to predict the CMB sky observed by a static observer who is displaced with respect to the center of a local underdense region. The distribution of matter as seen by this observer will be anisotropic. This will affect the observed distribution of CMB temperature and constrain the possible location of the observer. 
We model the local underdense region as a void region described by an LTB metric, embedded in a homogeneous and isotropic spacetime (the LTB metric, and the method described here, allow of course also for more general situations). 
The spherically symmetric LTB metric is given by
\beq \label{LTBmetric2}
\dd s^2 = -\dd t^2 + \frac{R'(r,t)^2}{1+\beta(r)}\dd r^2 + R(r,t)^2 \dd \Omega^2\,,
\eeq
where $R(r, t)$ is a position-dependent scale factor and $\beta(r)$ is related to the curvature. In the following, a dot denotes derivative with respect to physical time and a prime denotes derivative with respect to the radial coordinate. 

\begin{figure}
\centering
\includegraphics[scale=0.30]{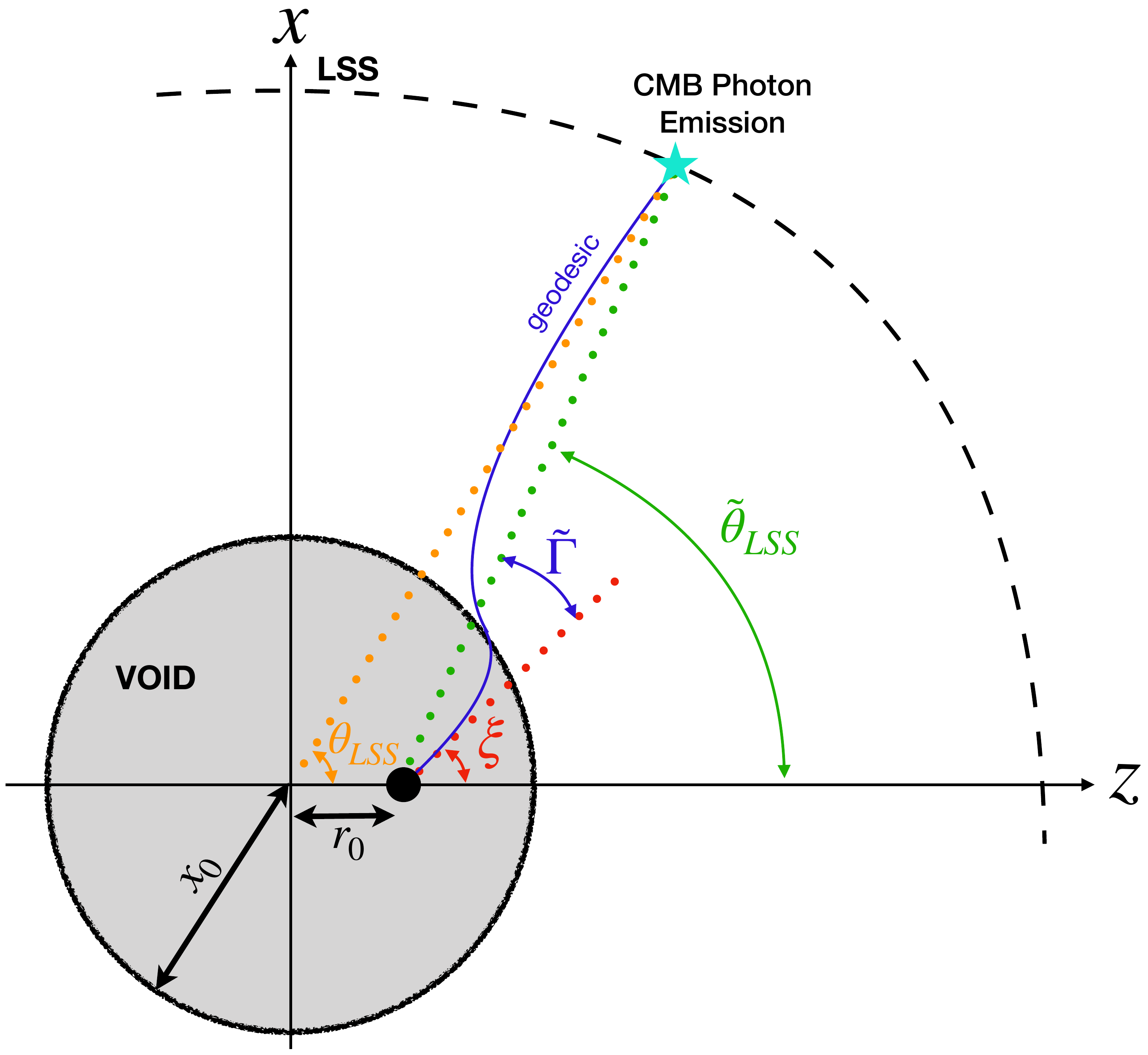}
\caption{ \small \label{model_schematic} A schematic illustration of the situation considered in this paper: In a locally underdense region (the void) that transitions to the `true' background at a radius of about $x_0$, the observer is located at the radial coordinate value of $r_0$ from the center of the void. Photons that reach the observer from a direction $\xi$ with respect to the void-center - observer axis follow a curved geodesic due to the presence of the void, which leads to a deflection angle $\tilde\Gamma$ between the emission location on the last-scattering surface and the arrival direction.}\end{figure}

\subsection{Background equations}
In a matter dominated universe the background evolution is described by the following set of equations,
\begin{align} \label{EinsteinEq1}
&H_{\perp}^2 + 2 H_{\perp} H_{\parallel} - \frac{\beta}{R^2} -\frac{\beta'}{RR'} = \kappa \rho\,,\\
&2R\ddot{R} + \dot{R}^2 = \beta\,, \label{EinsteinEq2}
\end{align}
where $H_{\perp} = \dot{R}/R$, $H_{\parallel}=\dot{R'}/R'$, $\kappa = 8\pi G$, and $\rho$ is the energy density of matter. Some alternative approaches and ideas regarding the LTB universe model are also given in Refs. \cite{GarciaBellido:2008nz, Biswas:2010xm, Enqvist:2007vb, Moffat:2005yx, Clarkson:2007pz, Nadathur:2010zm}.\\
Integrating Eq.\ (\ref{EinsteinEq2}) with respect to physical time, $t$, we get
\beq \label{EinsteinEq22}
H_{\perp}^2 = \frac{\alpha}{R^3} +\frac{\beta}{R^2}\,,
\eeq
where $\alpha$ is a function of $r$. This has the form of the Friedmann equation in a universe with matter and curvature. The matter density distribution as a function of radial coordinate, $r$, can be computed by differentiating Eq.\ (\ref{EinsteinEq22}) with respect to $r$ 
\beq \label{diffEinsteinEq22wrt_r}
2H_{\perp} H_{\parallel} = \frac{\beta'}{RR'} + \frac{\alpha'}{R'R^2} - \frac{\alpha}{R^3}\,,
\eeq
which can be used in Eq.\ (\ref{EinsteinEq1}) to obtain
\beq \label{kappa_rho}
\kappa \rho = \frac{\alpha'}{R'R^2}\,.
\eeq
Eq.\ (\ref{EinsteinEq22}) can be solved by introducing a conformal time parameter, $\eta$, by
\be
\sqrt{\beta}dt=R d\eta\,,
\ee
and we find 
\begin{subequations}
\beq \label{EinsteinSol1}
R(r,\eta)=\frac{\alpha}{2\beta}(\cosh\eta-1)+R_{\rm LSS} \Bigg(\cosh\eta + \sinh\eta\sqrt{\frac{\alpha}{\beta R_{\rm LSS}} + 1}\Bigg)\,,
\eeq
\beq \label{EinsteinSol2}
t(r,\eta)=\frac{\alpha}{2\beta^{3/2}}(\sinh\eta-\eta)+\frac{R_{\rm LSS}}{\sqrt{\beta}} \Bigg( \sinh\eta + (\cosh\eta-1)\sqrt{\frac{\alpha}{\beta R_{\rm LSS}} + 1}\Bigg)\,,
\eeq
\end{subequations}
where we have defined ${R_{\rm LSS}}(r)\equiv R(r,t=0)$, at some initial $t=0$ where we assume the last-scattering surface to be located.

Here we want to model a large scale underdense void where the spacetime is defined by the LTB metric, embedded in a nearly-flat FLRW universe outside this void. So $\alpha$ and $\beta$ are chosen in such a way to reproduce these asymptotic regimes, smoothly interpolating between them 
\begin{align} \label{alpha}
\alpha(r) &= H_{\perp,0}^2 r^3 \Bigg[\alpha_0 - \frac{\Delta\alpha}{2}\Bigg(1-\tanh{\frac{r-x_0}{2\Delta x}} \Bigg) \Bigg]\,,\\
\beta(r) &= H_{\perp,0}^2 r^2 \Bigg[\beta_0 - \frac{\Delta\beta}{2}\Bigg(1-\tanh{\frac{r-x_0}{2\Delta x}} \Bigg) \Bigg]\,, \label{beta}
\end{align}
where $H_{\perp,0}$ is the present value of the transverse Hubble parameter in the outer FLRW region, $\alpha_0$ and $\beta_0 = 1-\alpha_0$ are the relative (reduced) densities of matter and curvature in this region, $\Delta\alpha$ and $\Delta\beta$ specifies the difference between the matter density and curvature density, respectively, in the two regions, and $x_0$ and $\Delta x$ are the point and width of the transition from one region to another. We observe that in the far outer region $r\gg x_0$, one has a nearly-flat homogenous universe with 
\beq \label{+veFLRW_R}
R = \frac{1}{2}\frac{\alpha_0 r}{\beta_0} (\cosh{\eta}-1) \equiv r a(\eta)\,,
\eeq
\beq \label{+veFLRW_t}
t = \frac{1}{2H_0} \frac{\alpha_0}{\beta_0^{3/2}} (\sinh\eta - \eta) \equiv t(\eta)\,,
\eeq
where the function $a(\eta)$ represents the scale factor in this outer homogeneous region. It follows that the function $R_{\rm LSS}$ in Eq.\ (\ref{EinsteinSol1}) and (\ref{EinsteinSol2}) can be naturally defined as 
\beq \label{Rlss}
R_{\rm LSS} = a_* r\,,
\eeq
where $a_*$ is the scale factor at the time of last-scattering.

\subsection{Photon geodesics}

We follow \cite{Alnes:2006pf}. Photons follow trajectories determined by the geodesic equation 
\be
\frac{d^2 x^{\mu}}{d\lambda^2}+\Gamma^{\mu}_{\alpha \nu}\frac{dx^{\alpha}}{d\lambda}\frac{dx^{\nu}}{d\lambda}=0\,,
\ee
where $\Gamma^{\mu}_{\alpha \nu}$ is the Christoffel symbol and $\lambda$ is an affine parameter along the path of the photon. Due to axial symmetry, the photon paths must be independent of the azimuth angle $\phi$, which leaves three possible choices for the free index $\mu$. For $\mu=t$ we have 
\be\label{1}
\frac{du}{d\lambda}\equiv \frac{d^2t}{d\lambda^2}=-\frac{R'\dot{R}'}{1+\beta}\left(\frac{dr}{d\lambda}\right)^2-R\dot{R}\left(\frac{d\theta}{d\lambda}\right)^2\,.
\ee
Next, $\mu=r$ gives
\be
\frac{dp}{d\lambda}\equiv\frac{d^2 r}{d\lambda^2}=-\left(\frac{R''}{R'}+\frac{\beta'}{2+2\beta}\right)\left(\frac{dr}{d\lambda}\right)^2-\frac{2\dot{R}'}{R'} \frac{dr}{d\lambda}\frac{dt}{d\lambda}+\frac{R(1+\beta)}{R'}\left(\frac{d\theta}{d\lambda}\right)^2\,.
\ee
The equation for $\mu=\theta$ can be written as a conservation equation for the angular momentum $J$ as
\be\label{2}
\frac{d}{d\lambda}\left(R^2 \frac{d\theta}{d\lambda}\right)\equiv \frac{d}{d\lambda}J=0\,.
\ee
Finally, the null-geodesic condition gives the constraint equation
\be\label{null_constraint}
-\left(\frac{dt}{d\lambda}\right)^2+\frac{(R')^2}{1+\beta}\left(\frac{dr}{d\lambda}\right)^2+\frac{J^2}{R^2}=0\,.
\ee

We specify initial condition at the time $t_0$ when the photon arrives at the observer. We choose a static observer at $r=r_0$ and $\theta=0$ (i.e., along the polar axis at $r=r_0$ from the center). The spatial components (along the $r$, $\theta$ and $\phi$ directions) of the unit vector along the $z$-axis are
\be
v^i=\frac{\sqrt{1+\beta}}{R'}\left(1, 0, 0\right)\,. 
\ee
We choose to parametrize the photon path in such a way that $\lambda=0$ at the observer position $r(0)=r_0$ and time of observation $t(0)=t_0$. We also choose $dt/d\lambda|_{\lambda=0}=u_0=-1$. Inserting this choice in Eq.\ (\ref{null_constraint}) we find initial conditions for $dr/d\lambda=p$ 
\be
p_0=\cos\xi \frac{\sqrt{1+\beta}}{R'}\Big|_{\lambda=0}\,.
\ee
The spatial direction of observation is given by the tangent to the photon path at $t_0$, i.e.,
\be\label{vec}
n^i=-\left(p, J/R^2, 0\right)\Big|_{\lambda=0}\,, 
\ee
normalized in such a way that $g_{ij} n^i n^j=1$. The angle at which the photon is received is given by the inner product of $v^i$ and $n^i$
\be
\cos\xi=g_{ij} v^i n^j=-\frac{R' p}{\sqrt{1+\beta}} \Big|_{\lambda=0}\,.
\ee
This equation can be used to express the angular momentum as a function of the angle $\xi$ as
\be
J=J_0=R(r_0, t_0)\sin \xi\,.
\ee
We can then solve the set of equations (\ref{1}), (\ref{2}) and (\ref{null_constraint}) with initial conditions $r_0$, $t_0$, $p_0$ and $J_0$.\footnote{We observe that since we have $t=t(r, \eta)$, an equation for $\eta$ can be derived using
\be
\frac{d\eta}{d\lambda}=\frac{\partial \eta}{\partial t}\frac{d t}{d \lambda}+\frac{\partial \eta}{\partial r}\frac{d r}{d \lambda}\,.
\ee}
The redshift as a function of $\lambda$ for a photon hitting the observer today can be found integrating the following equation, see e.g. \cite{Alnes:2006pf}
\be\label{z_geod}
\frac{d \ln(1+z)}{d\lambda}=-u^{-1} \left[\frac{R'\dot{R}'}{1+\beta} p^2+\frac{\dot{R}}{R^3} J^2\right]\,, 
\ee
with initial condition $z(\lambda=0)=z_0=0$.

\subsection{Reference models}

To derive numerical predictions for various observables, we need to choose values for the model parameters. For simplicity, we follow here Refs. \cite{Alnes:2005rw, Alnes:2006pf}, which chose parameters such that the distances agree with the observations of SN-Ia for low redshifts and those of the cosmic microwave background for high redshifts. We emphasize however that fitting distance measurements is not our goal in this paper, and that we use these parameters only to illustrate the resulting effects. A more detailed and systematic study of different void sizes will be presented in a companion publication.

The density distribution parameters $\alpha$ and $\beta$ entering the expression (\ref{EinsteinEq22}) of the Hubble parameters are functions of the spatial coordinate $r$.
The universe outside the void is defined by nearly-flat FLRW dust dominated universe. So the relative matter density in this region can be set to $\alpha_0 = 1$. Consequently, the relative curvature density becomes $\beta_0 = 1-\alpha_0 = 0$ in this region. The curvature density decreases over the transition region by the same amount as the increase in matter density with increase in $r$. So the density contrast parameters are of opposite sign but same magnitude, i.e., $\Delta\alpha= -\Delta\beta$. It follows that the free parameters that remain are $\Delta\alpha$, $H_{\perp,0}$, $x_0$, and $\Delta x$ which are model dependent.
Note that, for practical reasons, we use $\alpha_0 = \Omega_{m,\text{out}} = 0.999$ instead of $\alpha_0 = 1$ so that $\beta_0 = 1 - \alpha_0$ does not vanishes completely in the outside homogeneous FLRW region, which would otherwise cause numerical errors.

Following \cite{Alnes:2005rw}, a good fit to the supernovae observations for low redshifts requires the Hubble parameter in the vicinity of the observer or inside the void today to be $h_{\rm in} = 0.65$. Similarly, a good fit to the CMB power spectrum for high redshifts requires the Hubble parameter well outside the void today to be $h_{\rm out} = 0.51$.
 The $h_{\rm in/out}$ are given by $H_{\perp,0, {\rm in/out}}= 100 \times h_{\rm in/out} \text{km}\cdot \text{s}^{-1}\text{Mpc}^{-1}$. 

\begin{figure}
\centering
\includegraphics[scale=0.4]{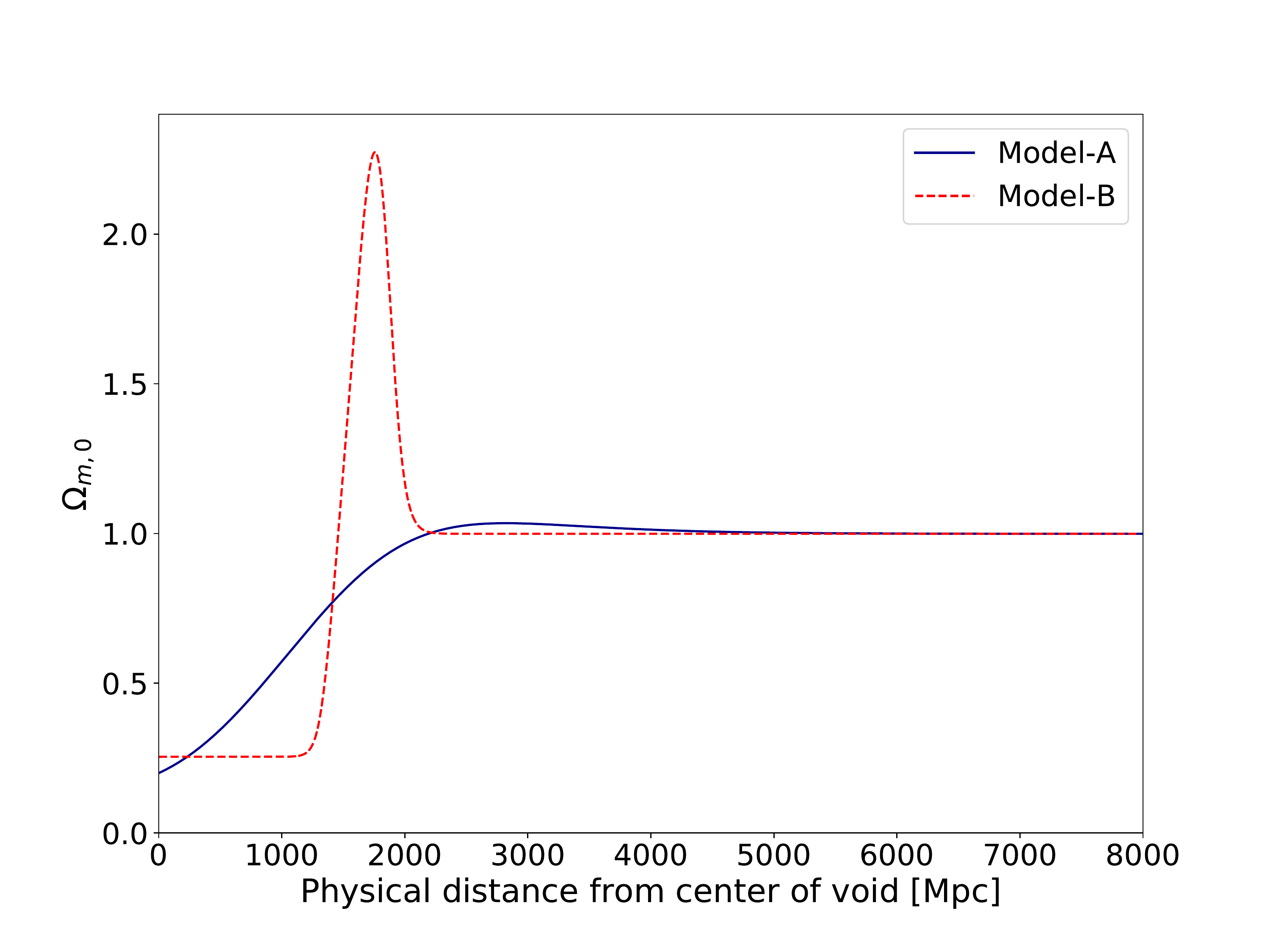}
\caption{\small Matter density distribution profiles at current time for the two models-A and -B defined earlier}
\label{LTB_matter_density_profiles}
\end{figure}

\begin{figure} %
    \centering
    \subfloat[LTB Model-A]{{\includegraphics[width=7.2cm]{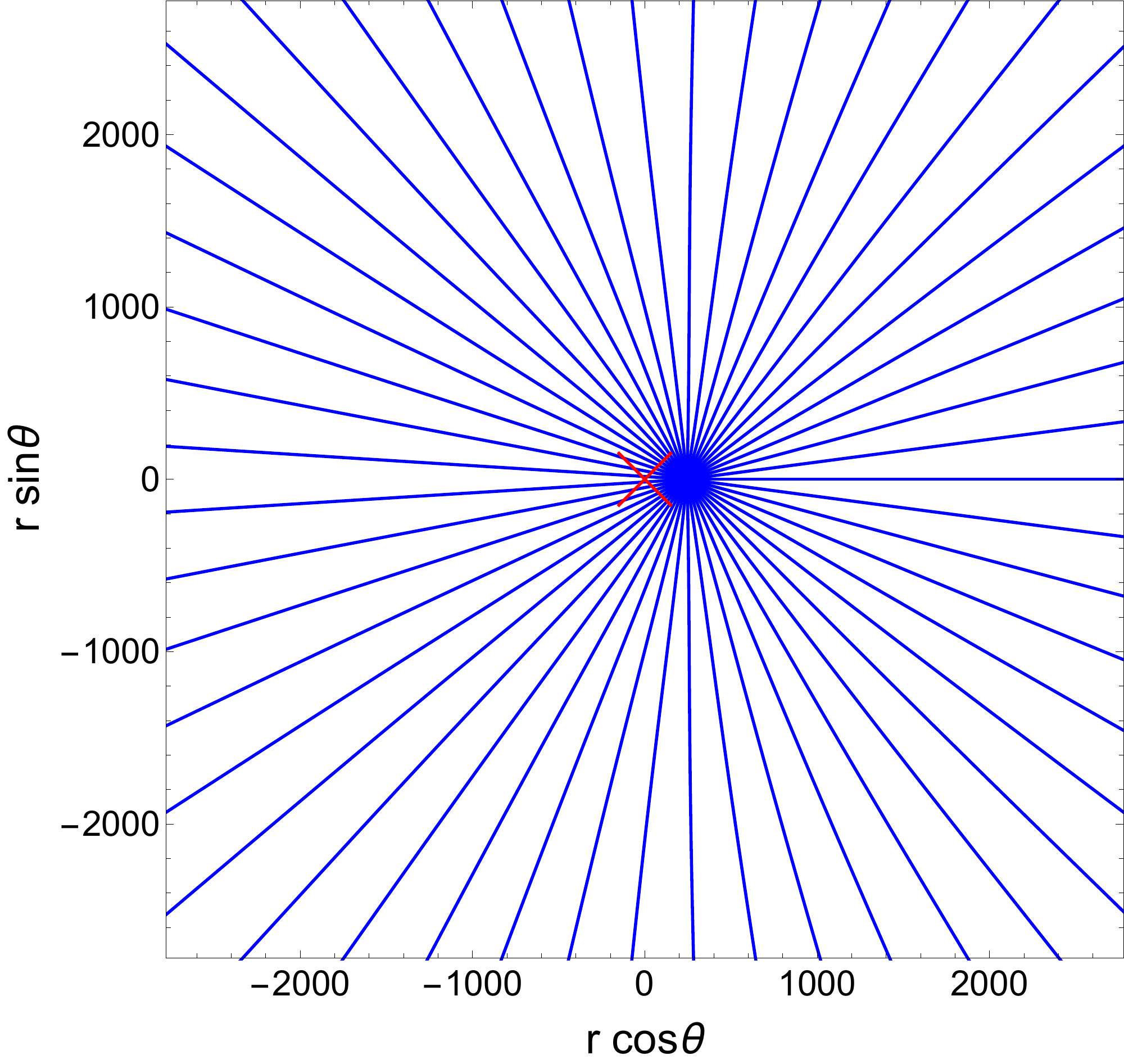} }}%
    \qquad
    \subfloat[LTB Model-B]{{\includegraphics[width=7.2cm]{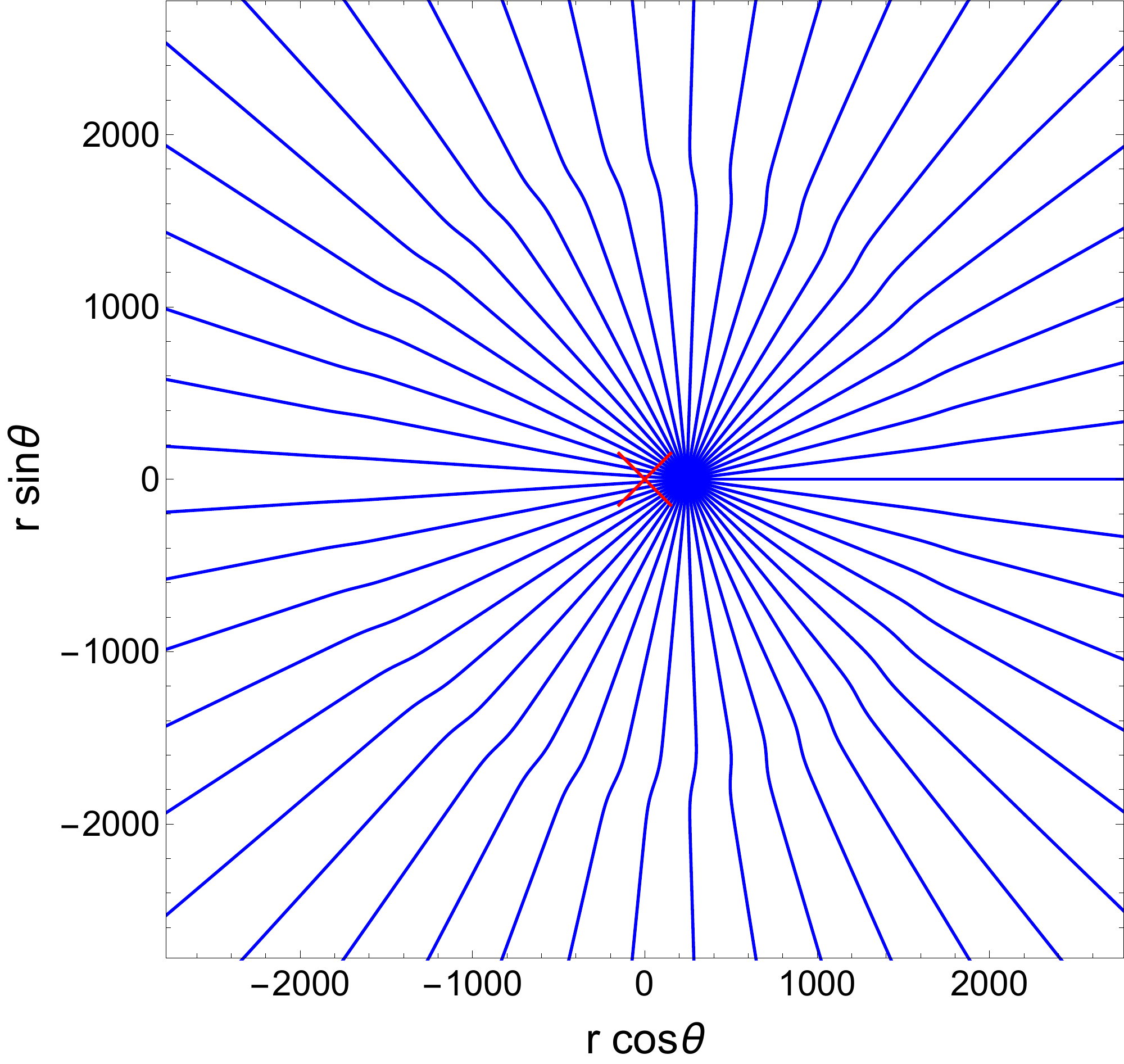} }}%
    \caption{\small Photon trajectories for observer at physical distance 200 Mpc today in models-A and -B with different matter density distribution profiles as shown in Fig. \ref{LTB_matter_density_profiles}}%
    \label{fig:trajAB} %
\end{figure}

For illustrative purposes, we focus on two specific models which we call model-A and model-B \cite{Alnes:2006pf, Alnes:2005rw},  whose parameters are listed in table \ref{paramval}. In Fig. \ref{LTB_matter_density_profiles} we show the matter density profile for each of these two models. In Fig.\,\ref{fig:trajAB} we show the photon trajectories for an observer at a physical distance 200 Mpc, for both models A and B. Not surprisingly, given the density profiles, the bending of the light rays due to the void is much sharper and more visible for void model-B. We shall however see later that the lensing effect is more pronounced in model-A.

\begin{table}[t]
\begin{center}

\begin{tabular}{l | c | c | r}
	\hline \hline	
	Parameter & Symbol & Model-A & Model-B \\
	\hline
	Relative matter density & $\alpha_0$ & 0.999 & 0.999 \\
	Relative curvature density & $\beta_0=1-\alpha_0$ & 0.001 & 0.001 \\
	Density contrast parameter & $\Delta\alpha = -\Delta\beta$ & 0.9 & 0.78  \\
	Transition point [Gpc] & $x_0$ & 1.450 &1.804 \\
	Transition width & $\Delta x / x_0$ & 0.40 & 0.03 \\
	Age of universe [Gyr] & $t_0$ & 12.8 &12.7 \\
	Relative density at the center & $\Omega_{m, {\rm in}}$ & 0.2 & 0.25\\
	Relative density outside void & $\Omega_{m, {\rm out}}$ & 0.999 & 0.999\\
	Hubble parameter at the center & $ h_{\rm in}$ & 0.65 & 0.63\\
	Hubble parameter outside void & $ h_{\rm out}$ & 0.51 & 0.51\\
	\hline \hline
\end{tabular} 
\caption{Parameter values used for calculation for both the models}
\label{paramval} 
\end{center}
\end{table}

\section{CMB angular power spectrum}\label{CMB}

Since space-time is no longer spherically symmetric around such an off-set observer, we expect her/him to measure additional anisotropies in the temperature, relative to those measured by an observer at the center. Following the approach of \cite{Cusin:2016kqx}, we expand the observed temperature anisotropy field up to first order in lensing-like displacement and radial modulation and we write  
\be\label{tailm}
\tilde{\Theta}({\bm{\bx_0}}\,,t_0\,,{\bm{\bn}})=\Theta({\bm{\bx_0}}\,,t_0\,,{\bm{\bn}})+\Theta^{NS}({\bm{\bx_0}}\,,t_0\,,{\bm{\bn}})+\Theta^{\varphi}({\bm{\bx_0}}\,,t_0\,,{\bm{\bn}})+\Theta^{d}({\bm{\bx_0}}\,,t_0\,,{\bm{\bn}})\,,
\ee
where we have explicitly indicated the dependence on the observer position ${\bm{\bx_0}}$ and reception time $t_0$. In Eq.\ (\ref{tailm}), $\Theta({\bm{\bx_0}}\,,t_0\,,{\bm{\bn}})$ is the zeroth order contribution from the primary anisotropies while $\Theta^{\varphi}$ and $\Theta^d$ are the lensing-like and radial modulation effects, linear in lensing-like deflection and in the radial modulation, respectively. These contributions vanish for a perfectly isotropic CMB sky without primary anisotropies. The contribution $\Theta^{NS}({\bm{\bx_0}}\,,t_0\,,{\bm{\bn}})$ is given by non-stochastic anisotropies induced by the displaced position of the observer in the apparent anisotropic spacetime. This is a purely geometry-induced effect, present even if primary anisotropies are vanishing. In the following, we investigate separately these contributions.

\subsection{Non-stochastic anisotropies}\label{induced_ns}

An observer displaced in the void would measure anisotropies even when the temperature at the last-scattering is isotropic. These anisotropies are what we call \emph{non-stochastic anisotropies}: they are a purely geometrical effect, only due to the propagation of photons in the apparent anisotropic spacetime.

Assuming the CMB temperature to be isotropic at the last-scattering surface, the CMB temperature apparent to the observer today is given by
\begin{equation} \label{apparent_temp}
T(\bn)=\frac{T_*}{1+z(\bn)}
\end{equation}
where $T_*$ is the average temperature at the last-scattering surface, and $z(\bn)$ is the redshift of the last-scattering surface in the direction $\bn$. The apparent average temperature $\hat{T}$ at observer's position today is then given by
\begin{equation} \label{avgtemp_obs}
\hat{T} \equiv \frac{1}{4\pi} \int \dd\Omega \ T(\bn)\,. 
\end{equation}
The average redshift to the last-scattering surface can be obtained by combining Eq.\ (\ref{apparent_temp}) and Eq.\ (\ref{avgtemp_obs}) as
\begin{equation} \label{zavg}
1+z_* \equiv \frac{T_*}{\hat{T}}\,. 
\end{equation}
We can then write Eq.\ (\ref{apparent_temp}) as
\be
T(\bn)=\frac{T_*}{1+z_*+\delta z(\bn)}\equiv \frac{\hat{T}}{1+\delta(\bn)}\,,
\ee
where we have defined
\be
\delta(\bn)\equiv\frac{\delta z}{1+z^*}\,.
\ee
The observer measures the non-stochastic relative temperature variation 
\begin{equation} \label{rel_temp_var}
\Theta^{NS}(\bn) \equiv \frac{\Delta T}{\hat{T}} = \frac{T(\bn) - \hat{T}}{\hat{T}} =-\frac{\delta(\bn)}{1+\delta(\bn)}\simeq -\delta(\bn)\,.\end{equation}

\subsection{Secondary anisotropies}

The right way to proceed to take into account the effects of radial modulation is to write the CMB temperature field on the sky as the projection of sources $S$ which contribute in an optically thin regime, as done in \cite{Cusin:2016kqx}. By doing this, one can verify that the contribution of the radial modulation to the temperature anisotropy field is subdominant with respect to the one coming from lensing-like deflection, see e.g. \cite{Cusin:2016kqx} for details.\footnote{Indeed, the lensing depends on the angular gradient of the lensing potential and its observable consequences are weighted by a factor of order $\ell$. This has the effect of increasing the magnitude of the effect and shifting it to higher multipoles.} From now on we therefore focus only on the contribution of secondary anisotropies coming from lensing.

\begin{figure}
\centering
\includegraphics[scale=0.40]{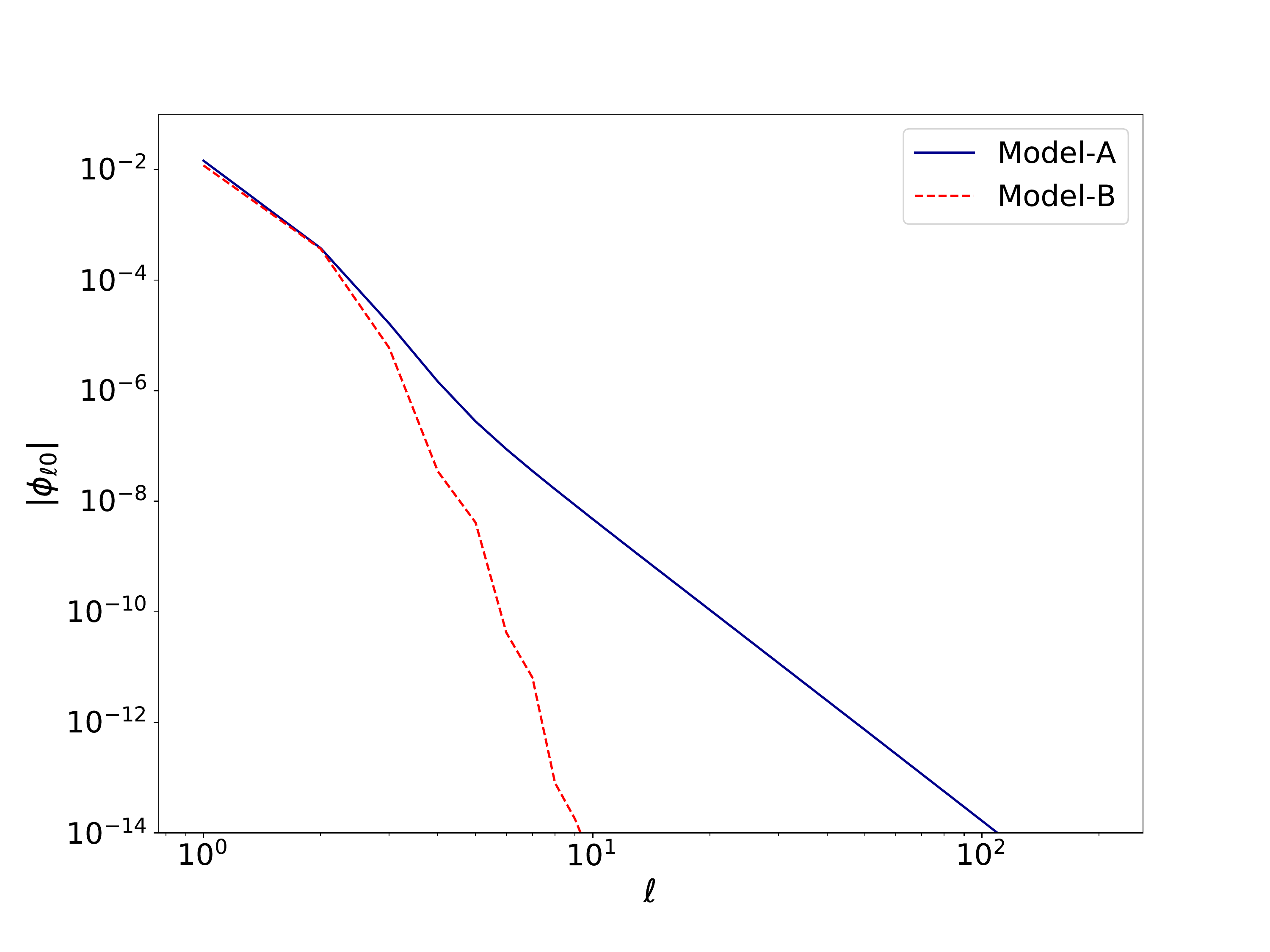}
\caption{\small \label{phil0_modelA} Multipoles of the lensing potential, $|\varphi_{\ell 0}|$, for LTB model-A and model-B}
\end{figure}

The lensing-like displacement as seen by the off-center observer is defined as the difference between the direction of the photon emission point at the last-scattering surface (the end-point of the photon geodesic) and the direction of observation (the angle of photon incidence at observer position).  Choosing the coordinate system as in Fig. \ref{model_schematic}, we define
\beq \label{lens_Gamma}
\tilde{\Gamma} \equiv \tilde{\theta}_{\text{LSS}} - \xi
\eeq
where with a tilde we denote quantities defined in the coordinate system centered at the observer position. The angle  $\tilde{\theta}_{\text{LSS}}$ is related to quantities defined in the centered reference frame through 
\beq \label{thtil}
\tilde{\theta}_{\text{LSS}} = \tan^{-1}\bigg({\frac{r_{\text{LSS}} \sin{\theta_{\text{LSS}}}}{r_{\text{LSS}} \cos{\theta_{\text{LSS}}} - r_0}}\bigg)\,,
\eeq
where $\theta_{\text{LSS}}$ and $r_{\text{LSS}}$ can be found by solving the geodesic equation. Since in the specific geometry chosen, lensing-like deflection has only gradient modes (for an extensive explanation, see appendix D1 of Ref.\,\cite{Cusin:2016kqx} that also applies here) a lensing potential can be introduced as $\Gamma_a = \nabla_a \varphi$ and the lensing-like displacement can be decomposed as 
\beq \label{phil01}
\tilde{\boldsymbol{\Gamma}} = - \sum_{\ell} \varphi_{\ell 0} \sqrt{\ell (\ell +1)} Y_{\ell 0} \boldsymbol{e}_{\theta}\,.
\eeq
The calculation of $\varphi_{\ell 0}$ is described in appendix D.1 of \cite{Cusin:2016kqx} with the final result given by
\beq \label{phil02}
\varphi_{\ell 0 } =
\frac{2 \pi}{\ell(\ell+1)}\sqrt{\frac{2\ell+1}{4\pi}}\int_{-1}^{1}\partial_{\xi}
P_\ell(\cos \xi) \tilde{\Gamma}(\xi) \dd\cos{\xi}\,.
\eeq
In Fig.\ \ref{phil0_modelA} we show $|\varphi_{\ell 0}|$ for models A and B. The function resulting from Eq.\ \ref{phil02} has a dependence on $(-1)^{\ell +1}$. The dipole and quadrupole of the two models is similar, but the higher multipoles of model B decrease very quickly. Those of model A scale like a power law, $\propto (-1)^{\ell +1} \ell^{-5.5}$, for $\ell \gtrsim 10$.

Until now we have considered a system of coordinates such that the azimuth was aligned with  ${\bf{e}}_z$, where ${\bf{e}}_z$ denotes the direction joining the center of the void to the observer.  To generalize our analysis, we can consider a rotated coordinate frame.  The rotation is described by a $\text{SO}(3)$ matrix $R_1$ characterized by its Euler angles $(\phi_1, \theta_1, 0)$.  In the new coordinate frame the direction observer-hole is described by the unit vector ${\bf{n}}_1= R_1 {\bf{e}}_z$. A direction described by a unit vector ${\bf{n}}$ in the old reference frame,  is rotated to $R_{1}^{-1} {\bf{n}}$ in the new one.  In this reference frame the lensing potential can be expanded as
\be\label{pot2}
\varphi_{\text{new}}({\bf{n}})=\sum_{\ell m} \varphi_{\ell m} Y_{\ell m} ({\bf{n}})\,,
\ee
with 
\be\label{mess}
\varphi_{\ell m}=\varphi_{\ell 0} \sqrt{\frac{ 4\pi}{2\ell +1}} Y_{\ell m}^*({\bf{n}_1})\,. 
\ee
The change to the rotated coordinate system does not change the results, however, and for the Fisher matrix analysis we will continue to use the preferred reference frame in which only the $m=0$ component of the lensing potential is non-zero.

\begin{figure}
\centering
\includegraphics[scale=0.40]{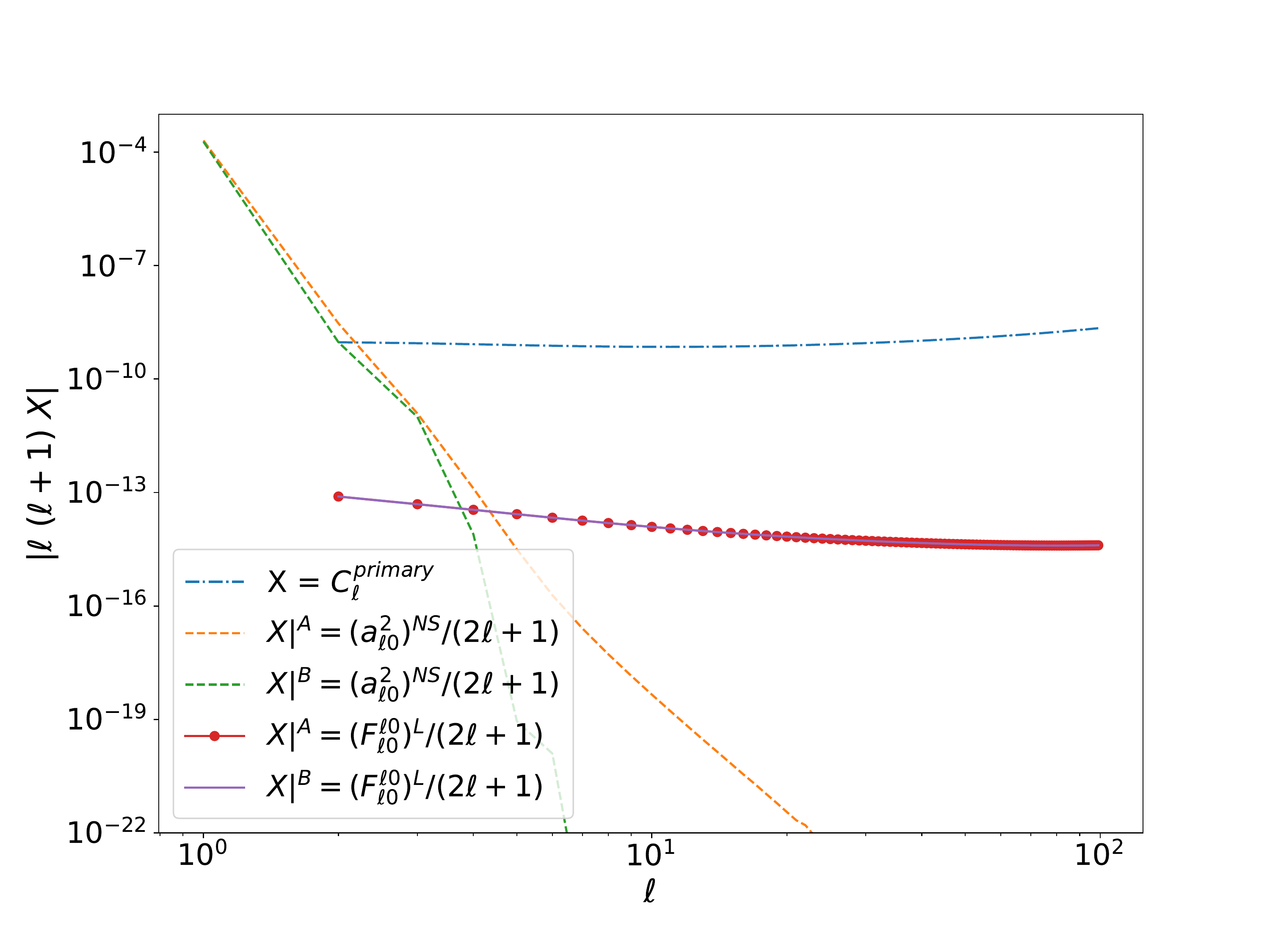}
\caption{\small \label{test_aniso_contri} Different contributions to the power spectrum: primary anisotropies, non-stochastic ones and lensing induced ones for LTB model-A and model-B.}
\end{figure}

\section{Correlation functions}\label{correlation}

The CMB sky measured by the off-center observer in the large scale void is not statistically isotropic. The correlation function of the observed temperature anisotropy (indicated by tilde) can be defined as 
\beq \label{correl_def}
\tilde{C}(\boldsymbol{n_1},\boldsymbol{n_2}) \equiv \langle \tilde{\Theta}(\boldsymbol{n_1}) \tilde{\Theta}(\boldsymbol{n_2})\rangle
\eeq
where $\boldsymbol{n_1}$ and $\boldsymbol{n_2}$ are the direction unit vectors in which the temperature at the last-scattering surface is measured. 

For the CMB analysis, we are interested in the two-point correlation function in harmonic space. We expand the temperature field as usual in spherical harmonics,
\beq
\tilde{\Theta}(\boldsymbol{n}) = \sum_{\ell m} \tilde{\Theta}_{\ell m} Y_{\ell m}(\boldsymbol{n}) \, ,
\eeq
and the two-point correlation function in harmonic space is then in general
\beq \label{2ptcorr}
F_{\ell m}^{\ell' m'} \equiv \langle \tilde{\Theta}_{\ell m} \tilde{\Theta}_{\ell'm'}^*\rangle \, .
\eeq
For a statistically isotropic random field on the sphere, the two-point correlation function would be diagonal, i.e.\ proportional to $\delta_{\ell \ell'} \delta_{m m'}$. In our anisotropic case however we expect off-diagonal contributions to $F$.
Since the secondary anisotropies have a pure geometrical origin, it is convenient to study the 2-point correlators for the lensing-like deflection and radial modulation separately and linearly sum them at the end. The 2-point function can be decomposed as \cite{Cusin:2016kqx}
\beq \label{flm_expansion}
F_{\ell m}^{L M} = C_{\ell}^{\Theta\Theta}\delta_{\ell\ell'}\delta_{mm'} + F_{\ell m}^{\ell' m'}|^{\varphi} + F_{\ell m}^{\ell' m'}|^d\, .
\eeq
In Eq.\ (\ref{flm_expansion}), the first term on the right hand side is the primary contribution  whereas the second and third terms denote the secondary contributions from lensing-like deflection and radial modulation, respectively. We stress that non-stochastic anisotropies are a pure geometric effect (they are not stochastic) hence they do not contribute to the correlation function. They can be treated as a mean value that in general depends on parameters such as void size/shape and observer position. 

As mentioned earlier, we focus on the 2-point correlator for the lensing-like deflection since the contribution from this effect is dominant with respect to that from the radial modulation effect. The second term on the right hand side of Eq.\ (\ref{flm_expansion}) can be explicitly written as
\beq \label{phi_corr_explicit}
F_{\ell m}^{\ell' m'}|^{\varphi} \equiv \langle \tilde{\Theta}^{\varphi}_{\ell m} \tilde{\Theta}^*_{\ell' m'}\rangle + \langle \tilde{\Theta}_{\ell m} \tilde{\Theta}^{*\varphi}_{\ell' m'}\rangle \,.
\eeq
As shown in Sect. 3.2 of \cite{Cusin:2016kqx}, the final result for 2-point temperature anisotropy correlation function is given by
\beq \label{Flm_final}
F_{\ell m}^{\ell' m'}|^{\varphi} = \varphi_{\ell_1 m-m'} \ \mathcal{C}^{m \ m' \ m-m'}_{\ell \ \ell' \ \ell_1} (\alpha_+ C^{\Theta\Theta}_{\ell'} + \alpha_- C^{\Theta\Theta}_{\ell})
\eeq
with 
\beq \label{alphapm} 
\alpha_{\pm} = \frac{1}{2} \big[ \ell_1(\ell_1+1) \pm (\ell'-\ell)(\ell'+\ell+1) \big].
\eeq
The summation over $\ell_1$ is understood, with $\ell_1$ going from $|\ell'-\ell|$ to $(\ell' + \ell)$ according to the triangle inequalities of Wigner 3-$j$ symbols in appendix H of \cite{Cusin:2016kqx}. The object $\mathcal{C}^{\ldots}_{\ldots}$ is called the Gaunt coefficient defined in the appendix. 
Because of the constraint on the upper indices in the Wigner 3-$j$ symbol that appears in $\mathcal{C}$, we set $M=m-m'$ in $\varphi_{\ell_1 M}$ in Eq.\ (\ref{Flm_final}). For our special choice of coordinates mentioned earlier, we have here $M=0$, i.e., $m=m'$. This also greatly reduces the number of coefficients that we have to consider while calculating the correlation function. $C_{\ell}^{\Theta\Theta}$ is the dimensionless primary contribution to the CMB temperature anisotropies calculated using the CLASS code \cite{Lesgourgues:2011re, Blas:2011rf}. We use the default values of parameters in the CLASS code to compute $C_{\ell}^{\Theta\Theta}$, as they reproduce well the observed CMB anisotropy spectrum. The most important parameter values are given in Table \ref{CLASSparam}. 

\begin{table}[t]
\begin{center}

\begin{tabular}{l | c | r}
	\hline \hline	
	Parameter & Symbol & Value \\
	\hline
	Hubble parameter [$\text{km} \cdot \text{Mpc}^{-1} \cdot \text{s}^{-1}$] & $H_0$ & 67.556 \\
	CMB temperature [K] & $T_{\text{CMB}}$ & 2.7255 \\
	Baryon density & $\Omega_b h^2$ & 0.022032 \\
	Cold dark matter density & $\Omega_{\rm CDM} h^2$ & 0.12038 \\
	Curvature density & $\Omega_K$ & 0 \\
	\hline \hline
\end{tabular} 
\caption{Parameter values used in CLASS code}
\label{CLASSparam} 
\end{center}
\end{table}

Figure \ref{test_aniso_contri} shows the different contributions to the CMB power spectrum, i.e., $\ell (\ell+1) F_{\ell m}^{\ell' m}$. The primary anisotropies are the same as those used in Eq.\ (\ref{Flm_final}). The spectrum of the non-stochastic anistotropies was computed with the help of a harmonic transform of the non-stochastic temperature variation (\ref{rel_temp_var}), while the lensing spectrum was obtained with the help of Eq.\ (\ref{Flm_final}). We divide the non-stochastic and lensing contributions by $2\ell + 1$ as they are only non-zero for $m=0$.  The lensing spectrum is much smaller than the primary anisotropies. The combination of void size and observer position used here leads to a large non-stochastic dipole and to an non-stochastic quadrupole that is comparable with the primary quadrupole, for a static observer. Realistically, the observer would be moving, which would change the lowest multipoles due to the boost contribution discussed in Sect.\ \ref{sec:kinematic} below. The non-stochastic anisotropies decay rapidly at higher $\ell$ and can be safely neglected there.

\begin{figure}
\centering
\includegraphics[scale=0.40]{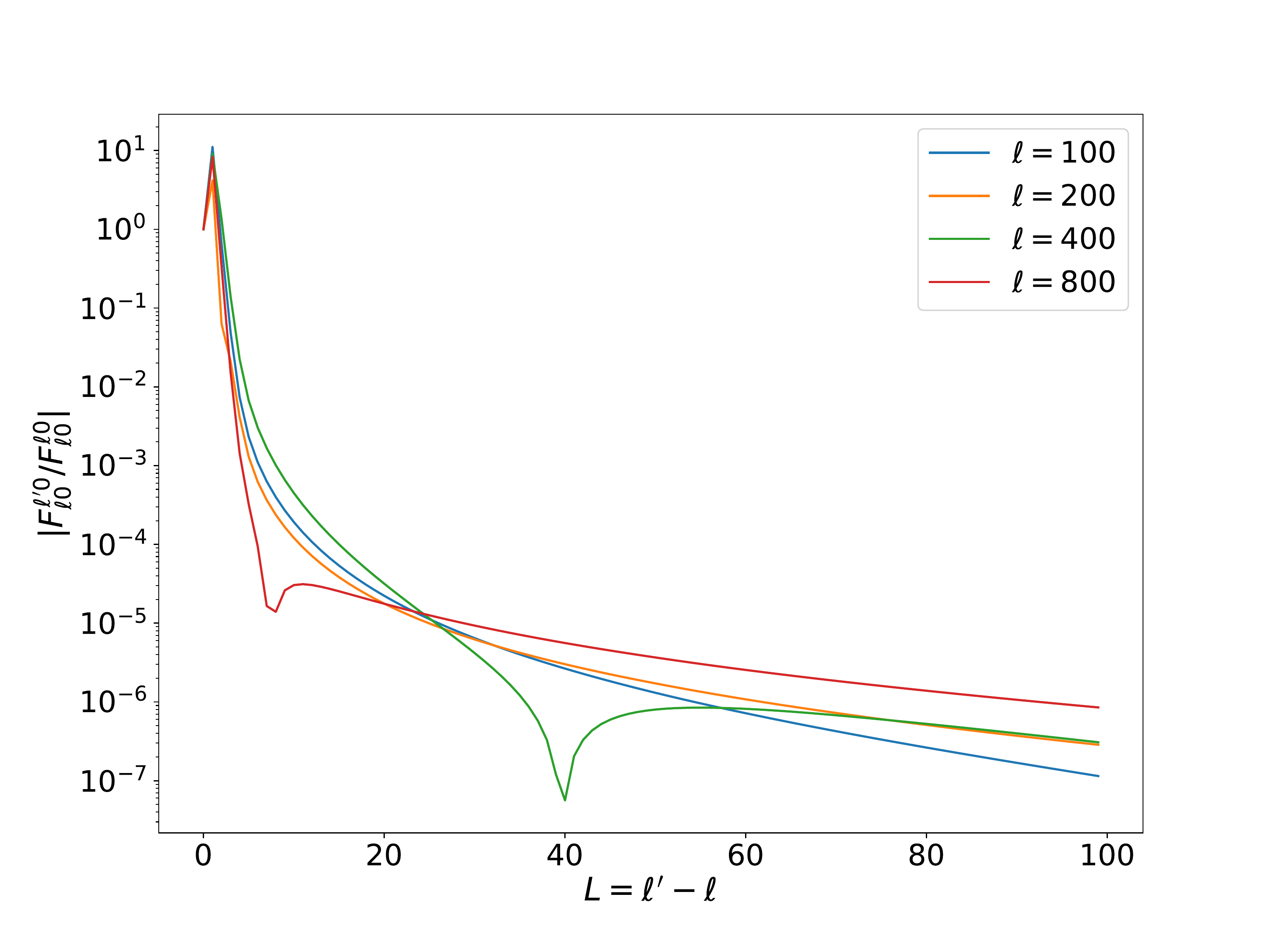}
\caption{\small \label{corr_m1}Normalized correlation matrix for different values of $\ell$ for LTB model-A}
\end{figure}

\begin{figure}
\centering
\includegraphics[scale=0.40]{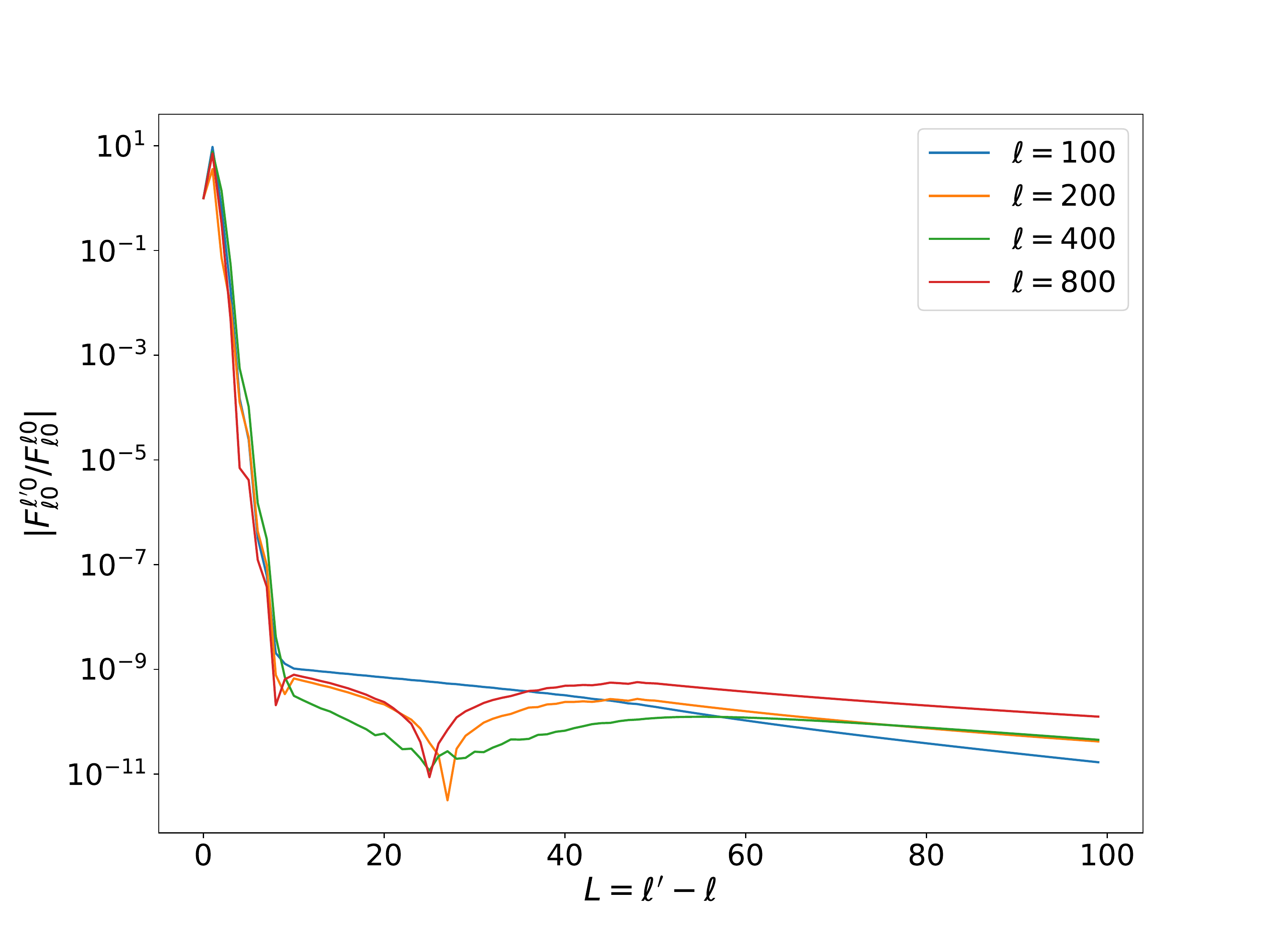}
\caption{\small \label{corr_m2}Normalized correlation matrix for different values of $\ell$ for LTB model-B}
\end{figure}

In Figs. \ref{corr_m1} and \ref{corr_m2}, we show the structure of normalized (to the diagonal) 2-point correlation function with $m=m'=0$ with $L=\ell' - \ell$ on the x-axis indicating the off-diagonal elements of the function (which appear in our case due to statistical anisotropy). An important observation from these figures is how the function decreases rapidly in amplitude for higher values of $L$. This helps us provide an estimate of the minimum number of off-diagonal elements that we can include in the numerical calculation of the Signal-to-Noise ratio (Sect.\ \ref{results}) after which (i.e., including more off-diagonal elements) the result remains effectively the same. We see also here that the off-diagonal correlations of model B decay faster than those of model A. \\

 \subsection{Inclusion of kinematic effects\label{sec:kinematic}}
 
 Up to now we have considered the case of an observer static (with respect to the CMB rest frame) inside the void. Here we explore the effect of a peculiar motion on the correlation function. We assume that kinematic effects  are adding up to lensing induced effects, i.e., we neglect effects  proportional to the product of lensing potential and peculiar motion, which we treat as second order. In other words, we neglect here the presence of the void and we review the effect of a peculiar motion on primary anisotropies, see also \cite{Cusin:2016kqx} where the comparison between lensing induced and kinematic effects on the correlation function was discussed for the first time.

We consider two observers of a pure FLRW universe: the first one comoving with the CMB rest frame and the second one in motion with respect to the first. We relate CMB correlation functions in the CMB rest frame $\mathcal{S}'$ with the ones in the moving observer frame $\mathcal{S}$. There are three different effects on the CMB sky map due to the motion of the observer:  (a) creation of anisotropies from the monopole (b) a modulation of intensity/Stokes parameters and (c) an aberration in the direction ${\bm{n}}$ of incoming photons which leads to a remapping of the intensity map/Stokes parameters on the sky. Explicitly, the energy of the photons changes as,
\be\label{Eshift}
E(\bn)=\frac{\sqrt{1-\beta^2}}{1- \bn\cdot \bv}E'(\bn')\,,
\ee
where $\bv=\beta \hat{\bv}$ is the relative velocity of the two frames. The directions of observations are related by
\be\label{aberration}
\bn \cdot \hat{\bv}=\frac{\bn'\cdot \hat{\bv}+\beta}{1+\bn'\cdot\bv}\,.
\ee
Temperature anisotropies seen by the observer in $\mathcal{S}$ can be divided into two categories: modulation of the monopole (intrinsic, non-stochastic component) and modulation and aberration of temperature anisotropies in $\mathcal{S}$ (stochastic anisotropies). Explicitely
\be
\Theta(\bn)=\Theta(\bn)^{NS}+\Theta(\bn)^S\,,
\ee
where the labels NS and S stay for non-stochastic and stochastic anisotropies, respectively. The temperature shift associated to the energy shift (\ref{Eshift}) is just 
\be\label{Tshift}
T=\frac{\sqrt{1-\beta^2}}{1- \bn\cdot \bv}T'\,. 
\ee
Averaging this object over directions (integrating over angles) we find
\be
\bar{T}=\frac{1}{4\pi}\int d^2\bn\, T(\bn)=\frac{\sqrt{1-\beta^2}}{\beta}\log\left(\frac{1+\beta}{1-\beta}\right)\frac{\bar{T}'}{2}\,,
\ee
and it is then straightforward to compute the relative anisotropies
\be
\Theta(\bn)^{NS}=\frac{2 \beta}{1-\bn\cdot \bv}\ln\left(\frac{1+\beta}{1-\beta}\right)^{-1}-1\,.
\ee
Introducing a series expansion in powers of $\beta$ it is easy to verify that
\be
\Theta(\bn)^{NS}=\sum_{\ell} a_{\ell} P_{\ell}(\bn\cdot \hat{\bv})\,,
\ee
with $a_{\ell}\propto \beta^{\ell}$. Stochastic anisotropies are given by a modulation and aberration of the anisotropies in $\mathcal{S}$. It follows from what we computed above for the non-stochastic case, that stochastic anisotropies are given by 
\be
\Theta(\bn)^S=\frac{2 \beta}{1-\bn\cdot \bv}\ln\left(\frac{1+\beta}{1-\beta}\right)^{-1}\Theta'(\bn'(\bn))\,,
\ee
where the relation between $\bn$ and $\bn'$ is given by (\ref{aberration}). Expanding in powers of $\beta$, up to $\mathcal{O}(\beta^3)$
\be
\Theta(\bn)^S=(1+\zeta_{\beta}+\zeta_{\beta^2})\left[\Theta'(\bn)-\left(\nabla_a \zeta_{\beta}+\frac{1}{4} \nabla_a\zeta_{\beta^2}\right)\nabla^a \Theta'(\bn)+\frac{1}{2} \nabla^a \nabla^b \Theta'(\bn)\nabla_a \zeta_{\beta}\nabla_b \zeta_{\beta}+\dots \right]\,,
\ee
where to simplify the notation we have introduced two potentials, linear in $\beta$
\be
\zeta_{\beta}(\bn)=\beta \bn\cdot \hat{\bv}\,,
\ee
and quadratic in $\beta$ 
\be
\zeta_{\beta^2}(\bn)=\beta^2\left( (\bn\cdot \hat{\bv})^2-\frac{1}{3}\right)\,,
\ee
Secondary anisotropies are stochastic and we can compute the correlation matrix, following  \cite{Cusin:2016kqx}. We write
\be
\Theta(\bn)=\sum_{\ell m} Y_{\ell m}(\bn) \Theta_{\ell m}\,, \quad\Theta'(\bn)=\sum_{\ell m} Y_{\ell m}(\bn) \Theta'_{\ell m}\,,
\ee
and for the potentials
\be
\zeta_{\beta}(\bn)=\sum_{m} Y_{1 m}(\bn) \beta_{1 m}\,, \quad \zeta_{\beta^2}(\bn)=\sum_{m} Y_{2 m}(\bn)\beta_{2 m} \,.
\ee
At linear order in $\beta$, one has  \cite{Cusin:2016kqx}
\be
( F_{\ell m}^{\ell' m'})|^{\beta}=\alpha_{\ell'\ell}\,\beta_{1 m-m'} \,\mathcal{C}_{\ell\,\,\ell'\,\,1}^{m\,\,m'\,\,m-m'}(C_{\ell}-C_{\ell'})\,,\label{v1}\qquad \alpha_{\ell'\ell}\equiv \frac{\ell'-\ell}{2}(\ell'+\ell+1)\,, 
\ee
We observe all the diagonal terms (i.e. $\ell=\ell'\,, m=m'$) of the correlation matrices $( F_{\ell m}^{\ell' m'})|^{\beta}$ are vanishing. Off-diagonal correlators are non-vanishing only for $\ell'=\ell\pm1$, i.e. we have only correlation among $\ell\leftrightarrow \ell\pm 1$ multipoles. This result can be extended: at a generic order $L$ of the perturbation expansion in $\beta$, only off-diagonal elements separated at most by $L$ are excited. For example, the correlation matrix at order $\beta^2$ reads
\begin{align}
( F_{\ell m}^{\ell' m'})|^{\beta^2}=&-\frac{1}{4} \beta_{2 m-m'} \mathcal{C}_{\ell\,\,2\,\, \ell'}^{m\,\,(m-m')\,\,m'} \left(\alpha_{\ell'\ell}+1\right)C_{\ell'}\nn\\
&+\sum_{\ell'' m''}\beta_{1 m-m''}\beta_{1 m'-m''}\mathcal{C}_{\ell\,\,1\,\, \ell''}^{m\,\, (m-m'')\,\, m''} \mathcal{C}_{\ell'\,\,1\,\, \ell''}^{m'\,\, (m'-m'')\,\, m''}\alpha_{\ell''\ell}\alpha_{\ell''\ell'}C_{\ell''}\nn\\
&-\sum_{M}\beta_{1 M}\beta_{1 (m-M-m')} \left(\mathcal{Q}_{\ell\,\,1\,\,\ell'\,\,1}^{m\,\,M\,\,m'\,\,(m-M-m')}-\frac{1}{2}\mathcal{R}_{\ell\,\,1\,\,\ell'\,\,1}^{m\,\,M\,\,m'\,\,(m-M-m')}\right)C_{\ell'}\nn\\
&+(\ell\leftrightarrow \ell')\,,
\end{align}
where the symbols $\mathcal{C}_{...}$, $\mathcal{I}_{...}$ $\mathcal{Q}_{...}$ and $\mathcal{R}_{...}$ are a combination of Wigner symbols, defined in the appendix.  We see that at second order in $\beta$ only off-diagonal elements up to a separation of $2$ are present as a result of the selection rules coming from the triangular  conditions of the Clebsh-Gordan coefficients. As already mentioned, this result holds at any order in $\beta^N$: only multiples separated by $\pm N$ are present at that order. 
The effect of a boost on anisotropies is therefore equivalent of having a lensing-like potential $\phi$ whose multipoles are related by $\phi_{\ell+1}/\phi_{\ell}\sim \beta$.

As $\beta \approx 10^{-3}$, this scaling implies that the effective `lensing pontential' of the boost decreases by nearly 3 orders of magnitude per $\ell$, which is much faster than the scaling of model A, and even of model B, see Fig.\ \ref{phil0_modelA}. We can therefore `boost away' only the dipole contribution to the `void lensing' $\phi_{\ell 0}$, the remaining multipoles will not be significantly affected.


\section{Fisher forecasts}\label{results}

The likelihood for the $\tilde{\Theta}_{\ell m}$ is of the standard multivariate Gaussian form
\be
\ln \mathcal{L}=-\frac{1}{2}\left[\sum_{\ell m}\sum_{\ell' m'} \left( \tilde{\Theta}_{\ell m}^* - \tilde{\Theta}_{\ell m}^{NS*}\right) \left(F^{-1}\right)_{\ell  m}^{\ell' m'} \left(\tilde{\Theta}_{\ell' m'} - \tilde{\Theta}_{\ell' m'}^{NS} \right) +\ln \det\left(F\right)_{\ell m}^{\ell' m'}\right]+\text{const}\,. 
\ee
The non-stochastic anisotropies $\tilde{\Theta}_{\ell' m'}^{NS}$ are not inherently random, but are given by the geometry of the void and observer. For this reason we treat them as a mean value, that in general depends on parameters like void size/shape and observer position. The covariance matrix is given by the two-point correlation function (\ref{2ptcorr}). It will also depend on the void/observer geometry, and also on the cosmological parameters.

To simplify our notation, we introduce the following matrix form
\be
F_{\mu\nu}\equiv F_{\ell m}^{\ell' m'}\,,
\ee
where the first index $\mu\equiv (\ell, m)$ while the second $\nu\equiv (\ell', m')$. The theoretical covariance matrix $F^{-1}_{\mu\nu}$ depends in general on cosmological and geometrical parameters $\lambda_{A}$ and the uncertainty with which we can recover these parameters is given by the Fisher matrix 
\be
\mathcal{F}_{A B}=\Big< -\frac{\partial^2 \ln \mathcal{L}}{\partial \lambda_{A}\partial \lambda_{B}}\Big>\,,
\ee
which can be written more explicitly as
\be
\mathcal{F}_{AB}=\frac{1}{2} \sum_{\mu\nu\alpha\sigma} \left[(F^{-1})_{\mu\nu}\partial_{A} F_{\nu\sigma} (F^{-1})_{\sigma \alpha} \partial_{B} F_{\alpha \mu}\right]
+ \sum_{\mu\nu} \partial_{A} \tilde{\Theta}_{\mu}^{NS*} (F^{-1})_{\mu\nu} \partial_{B} \tilde{\Theta}_{\nu}^{NS} \,. 
\ee
In full generality, the theoretical correlation function can be split into a diagonal contribution from primary anisotropy and an off-diagonal one from lensing and from a boost, as
\be\label{off}
F_{\mu\nu}=C_{\mu} \delta_{\mu\nu}+A_V\delta C_{\mu\nu}+A_{\beta} \delta D_{\mu\nu}\,,
\ee
where $A_V$ and $A_{\beta}$ are two artificially introduced amplitudes of the lensing and boost contributions, respectively. We want to focus on the lensing part and we forecast the precision with which we can measure $A_V$. To this goal, we keep all cosmological and void/observer parameters fixed (parameter degeneracies can increase the error bars on $A_V$). In this case the Fisher matrix has only one element 
\be
\mathcal{F}_{A_V A_V}=\sigma_{A_V}^{-2}=\frac{1}{2}\sum_{\mu}T_{\mu}\,,
\ee
with
\be\label{Tmu}
T_{\mu}\equiv \sum_{\nu\sigma\rho}\delta C_{\mu\nu}(F^{-1})_{\nu\sigma}\delta C_{\sigma\rho}(F^{-1})_{\rho\mu}\,.
\ee
The non-stochastic anisotropies, $\Theta_\mu^{NS}$, do not contribute here as they do not depend on $A_V$. A more realistic analysis would vary the void/observer parameters, but our goal here is to check whether void lensing is detectable in the most favourable setting. 

Since in Eq.\ (\ref{off}), we expect the off diagonal part to be strongly suppressed relative to the diagonal one, we use
\begin{align}
F_{\mu\nu}&=\sqrt{F_{\mu\mu}}\sqrt{F_{\nu\nu}}\left(\delta_{\mu\nu}+\frac{\delta C_{\mu\nu}}{\sqrt{F_{\mu\mu}}\sqrt{F_{\nu\nu}}}\right)\nn\\
&=\sqrt{C_{\mu}}\sqrt{C_{\nu}}\left(\delta_{\mu\nu}+\epsilon_{\mu\nu}\right)\,,
\end{align}
where we have used  $F_{\mu\mu}=C_{\mu}$. Then for the inverse
\be
(F^{-1})_{\mu\nu}\simeq \frac{\delta_{\mu\nu}}{C_{\mu}}-\frac{\delta C_{\mu\nu}}{C_{\mu}C_{\nu}}-\frac{\delta D_{\mu\nu}}{C_{\mu}C_{\nu}}\,,
\ee
which gives for (\ref{Tmu}) 
\be
T_{\mu}\approx  \frac{1}{C_{\mu}}\sum_{\nu}\frac{\delta C_{\mu\nu}\delta C_{\nu\mu}}{C_{\nu}}+\dots\,,
\ee
where terms of order $\sim(\delta C/C)^3$ and  $\sim(\delta C/C)^2(\delta D/C)$ have been neglected. In other words, we see that a boost would give sub-leading corrections to the Signal-to-Noise of the void lensing, hence we neglect it in this analysis. To understand how the different terms in the Fisher matrix contribute to the Signal-to-Noise, we compute the square root of the $\mu$-dependent term
\be
(S/N)_{\mu}\equiv \sqrt{\frac{T_{\mu}}{2}}=\frac{1}{\sqrt{C_{\mu}}}\left[\sum_{\nu}\frac{(\delta C)_{\mu\nu}(\delta C)_{\nu\mu}}{ 2 C_{\nu}}\right]^{1/2}\,.
\ee
In the usual notation and taking the square, for future convenience
\be \label{SNR2}
(S/N)^2_{\ell m}\equiv\frac{T_{\ell m}}{2}=\frac{1}{C_{\ell m}}\left[\sum_{\ell' m'}\frac{(\delta C)_{\ell m}^{\ell' m'}(\delta C)_{\ell' m'}^{\ell m}}{ 2 C_{\ell' m'}}\right]\,.
\ee
As a test of the method, we verified that we agree with the results of \cite{Amendola:2010ty} when we forecast the Signal-to-Noise ratio of Planck to measure our local velocity using Eq.\ (\ref{v1}).

For the completeness of our analysis, we show the difference between a cosmic variance limited experiment and a real experiment by obtaining the Signal-to-Noise results for both the models (A and B) while including the instrument noise of the Planck satellite.
For simplicity we follow \cite{Amendola:2010ty} and replace $C_{\ell}$ in Eq.\ (\ref{SNR2}) with the quantity
\be
C_{\ell}^N = \frac{1}{f_{\text{sky}}} (C_{\ell}+N_{\ell})\,,
\ee
where $N_{\ell}$ is given by
\be
N_{\ell} = \theta_0^2 \sigma_T^2 e^{\ell(\ell+1) \theta_0^2 /(8 \ln 2)}\,,
\ee
and $f_{\text{sky}}$ is the fraction of sky covered by the experiment. 
As in  \cite{Amendola:2010ty}, we consider the sensitivity of the best channel, $\theta_0 = 7'$, $\sigma_T = 2 \times 10^{-6}$, and $f_{\text{sky}}= 0.85$. This is in relatively good agreement with the measured performance of Planck, e.g.\ \cite{Aghanim:2019ame}.

\begin{figure} [h]
\centering
\includegraphics[scale=0.50]{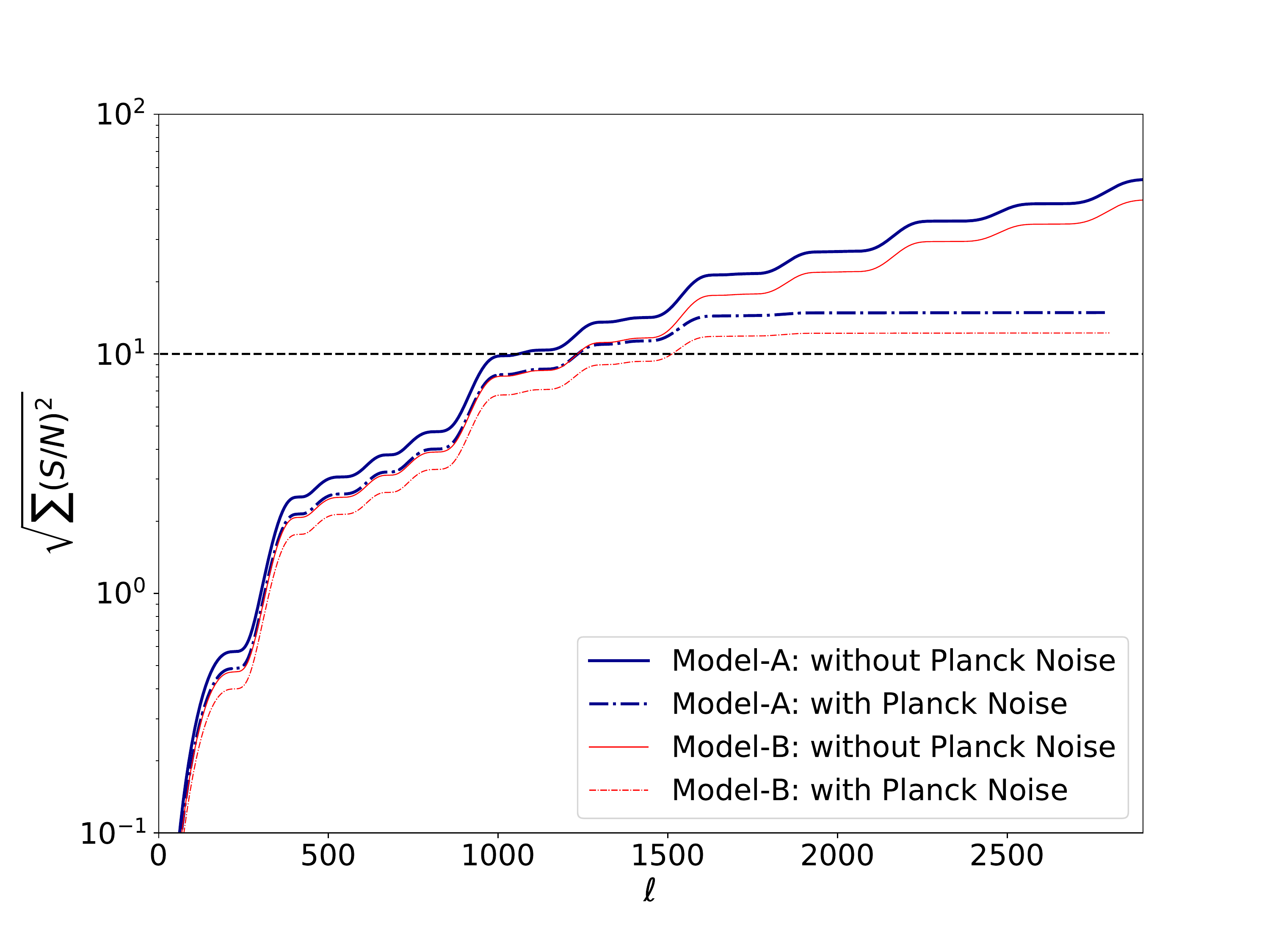}
\caption{\small \label{snr2cum_LTB_both_models} LTB: Results for cumulative Signal-to-Noise ratio with and without Planck Noise.}
\end{figure}

In Fig.\,\ref{snr2cum_LTB_both_models} we show the cumulative Signal-to-Noise for both model A and B, and an observer located at a physical distance of 200 Mpc with respect to the center of the void, as a function of $\ell_{\rm max}$. We see that the Planck signal levels off around $\ell \approx 1600$ where the noise starts to dominate. However, for both cases studied here, the Signal-to-Noise exceeds 10, i.e., both voids leave a detectable imprint in the Planck data. For purely cosmic variance limited surveys, the Signal-to-Noise ratio keeps increasing with resolution.

\section{Conclusions}

In this paper we have considered for the first time the structure of the temperature-temperature correlation matrix observed by an observer displaced with respect to the center of an underdense region described as an  LTB void model. We split the theoretical correlation function into a diagonal contribution from primary anisotropies and an off-diagonal one from lensing. We forecast the precision with which we can measure the amplitude of the lensing contribution. To this goal, we kept all cosmological and void/observer parameters fixed (parameter degeneracies can increase the error bars on the amplitude). We also discussed kinematic effects on the correlation matrix due to the peculiar velocity of the observer with respect to the CMB rest frame and we showed that in the Fisher forecast, the effect of additional boost-induced anisotropies manifests itself in sub-leading contributions to the Signal-to-Noise that we neglect in our approach. We applied the method to two large void models with different matter density profiles and we computed the Signal-to-Noise ratio for both cases. We find that these specific examples would leave a detectable signal in the Planck data, with a Signal-to-Noise ratio of over 10. A more extensive investigation is deferred to a companion paper dedicated to applying this framework to systematically to a large class of void models with different void sizes, observer positions and matter profiles where we also allow for a non-zero cosmological constant. In that publication we will also compare the results to the limiting cases of a empty spherical void embedded in a FLRW background where all the matter is concentrated in a black hole in the center, as in \cite{Cusin:2016kqx}, and the case of a void where the matter is concentrated in a thin shell at the void boundary.


\section*{Acknowledgements}
It is a pleasure to thank Joe Mohr for helpful discussions. We would also like to thank the referee for very useful comments. We acknowledge financial support from the Swiss National Science Foundation. GC acknowledges financial support from ERC Grant No:  693024 and Beecroft Trust. Part of the numerical calculations for this study were performed on the Baobab cluster of the University of Geneva.

\appendix

\section{Products of spherical harmonics and the Gaunt coefficient}\label{lm}

In this appendix we summarise some properties of integrals over products of three spherical harmonics, and associated quantities. It is based on appendix H of \cite{Cusin:2016kqx}. The main result is that the integral of three spin-weighted spherical harmonics can be written as
\begin{align}\label{1111}
\int d\Omega \,_{s_1}&Y_{\ell_1 m_1}\,_{s_2}Y_{\ell_2 m_2}\,_{s_3}Y_{\ell_3 m_3}=\nn\\
&=\sqrt{\frac{(2\ell_1+1)(2\ell_2+1)(2\ell_3+1)}{4\pi}}\left(
\begin{array}{ccc}
\ell_1&\ell_2&\ell_3\\
-s_1&-s_2&-s_3
\end{array}
\right)\left(
\begin{array}{ccc}
\ell_1&\ell_2&\ell_3\\
m_1&m_2&m_3
\end{array}
\right)\,.
\end{align}

The $3-j$ symbols that appear in this expression satisfy the following properties
\begin{align}\label{11}
\left(
\begin{array}{ccc}
\ell_1&\ell_2&\ell_3\\
m_1&m_2&m_3\\
\end{array}
\right)&=
\left(
\begin{array}{ccc}
\ell_2&\ell_3&\ell_1\\
m_2&m_3&m_1\\
\end{array}
\right)=
\left(
\begin{array}{ccc}
\ell_3&\ell_1&\ell_2\\
m_3&m_1&m_2\\
\end{array}
\right)\\
&=(-)^{\ell_1+\ell_2+\ell_3}
\left(
\begin{array}{ccc}
\ell_1&\ell_3&\ell_2\\
m_1&m_3&m_2\\
\end{array}
\right)\\
&=(-)^{\ell_1+\ell_2+\ell_3}
\left(
\begin{array}{ccc}
\ell_1&\ell_2&\ell_3\\
-m_1&-m_2&-m_3\\
\end{array}\label{111}
\right)\,.
\end{align}
Specifically, they are identically zero whenever any of the following conditions are violated
\be
m_1+m_2+m_3=0\,,\qquad |\ell_i-\ell_j|\leq \ell_k\leq \ell_i+\ell_j\,,\qquad \{i\,,j\}=\{1,2,3\}\,.
\ee

Some important quantities that are used in this paper are defined as
\begin{eqnarray}
\mathcal{C}^{m_1 m_2 m_3}_{\ell_1 \ell_2 \ell_3} &\equiv& \int \dd \Omega\,
Y^{\star}_{\ell_1 m_1} Y_{\ell_2 m_2} Y_{\ell_3 m_3}\,,\\
\mathcal{I}^{m_1 m_2 m_3}_{\ell_1 \ell_2 \ell_3} &\equiv& \int \dd \Omega\,
Y^{\star}_{\ell_1 m_1} \nabla^a Y_{\ell_2 m_2} \nabla_a Y_{\ell_3 m_3}\, .
\end{eqnarray}
The first quantity, $\mathcal{C}$, is effectively the Gaunt coefficient (up to the complex conjugation of the first spherical harmonic which leads to some sign changes). It is given by
\begin{align}
\mathcal{C}_{\ell_1 \ell_2 \ell_3}^{m_1 m_2 m_3} &=
(-1)^{m_1}\troisj{\ell_1}{\ell_2}{\ell_3}{-m_1}{m_2}{m_3}\,{\cal F}_{\ell_1\ell_2\ell_3}\,,\label{CC}\\
{\cal F}_{\ell \ell_1 \ell_2} &= \sqrt{\frac{(2\ell+1)(2
    \ell_1+1)(2 \ell_2+2)}{4\pi}}\troisj{\ell}{\ell_1}{\ell_2}{0}{0}{0}\, .
\end{align}
The second quantity is then
\begin{align}
 \mathcal{I}_{\ell_1 \ell_2 \ell_3}^{m_1 m_2 m_3} &= \frac{1}{2}\left[\ell_3
  (\ell_3+1)+\ell_2 (\ell_2+1)-\ell_1 (\ell_1+1)\right]\mathcal{C}_{\ell_1 \ell_2 \ell_3}^{m_1 m_2 m_3}\,,\label{EqItoC}\\
&=
(-1)^{m_1}\troisj{\ell_1}{\ell_2}{\ell_3}{-m_1}{m_2}{m_3}\,{F}_{\ell_1\ell_2\ell_3}\,,\\
F_{\ell \ell_1 \ell_2} &=
\frac{1}{2}[\ell_1(\ell_1+1)+\ell_2(\ell_2+1)-\ell(\ell+1)]  {\cal F}_{\ell \ell_1 \ell_2} \,.
\end{align}
It is easy to verify that
\be
 \mathcal{I}_{\ell_1 \ell_2 \ell_3}^{-m_1\, -m_2\, -m_3}=(-)^{\ell_1+\ell_2+\ell_3} \mathcal{I}_{\ell_1 \ell_2 \ell_3}^{m_1 m_2 m_3}\,.
 \ee
We also define
\begin{eqnarray}
\mathcal{Q}^{m_1 m_2 m_3 m_4}_{\ell_1 \ell_2 \ell_3 \ell_4} &\equiv& \int \dd \Omega\,
Y^{\star}_{\ell_1 m_1} Y_{\ell_2 m_2} \nabla^a Y_{\ell_3 m_3}\nabla_a Y_{\ell_4 m_4}\,,\\
\mathcal{R}^{m_1 m_2 m_3 m_4}_{\ell_1 \ell_2 \ell_3 \ell_4} &\equiv& \int \dd \Omega\,
Y^{\star}_{\ell_1 m_1} \nabla_aY_{\ell_2 m_2} \nabla^a\nabla^b Y_{\ell_3 m_3}\nabla_b Y_{\ell_4 m_4}\, .
\end{eqnarray}
For $\mathcal{Q}$ the following property can be verified:
\be
\mathcal{Q}^{m_1 m_2 m_3 m_4}_{\ell_1 \ell_2 \ell_3 \ell_4} =\sum_{\ell m} \mathcal{C}_{\ell_1\,\,\ell_2\,\,\ell}^{m_1\,\,m_2 \,\,m}\mathcal{I}_{\ell\,\,\ell_3\,\,\ell_4}^{m\,\,m_3 \,\,m_4}\,.
\ee


\newpage

\bibliographystyle{JHEP}
\bibliography{myrefs}

\end{document}